\documentclass[12pt]{iopart}
\usepackage{graphicx}
\pdfoutput=1

\providecommand{\boldsymbol}[1]{\mbox{\boldmath $#1$}}
\providecommand{\tabularnewline}{\\}

\usepackage{hyperref}
\usepackage{color}
\definecolor{red}{rgb}{0.9,0.3,0.3}

\makeatother

\begin{document}

\title{Coinage-metal capping effects on the spin-reorientation transitions
of Co/Ru(0001)}

\author{Farid El Gabaly$^{1,2}$, Kevin F. McCarty$^{3}$, Andreas K. Schmid$^{2}$,
and Juan de la Figuera$^{1,4}$}

\address{1 Centro de Microan\'{a}lisis de Materiales, Universidad Aut\'{o}noma de Madrid, Madrid 28049, Spain}
\address{2 Lawrence Berkeley National Laboratory, Berkeley 94720, USA}
\address{3 Sandia National Laboratories, Livermore, California 94550, USA}
\address{4 Instituto de Qu\'{\i}mica-F\'{\i}sica ``Rocasolano'', CSIC, Madrid 28006, Spain}

\author{M.~Carmen Mu\~noz$^{5}$, Laszlo Szunyogh$^{6}$, Peter Weinberger$^{7}$,
and Silvia Gallego$^{5}$}

\address{5 Instituto de Ciencia de Materiales de Madrid, CSIC, Madrid 28049, Spain}
\address{6 Department of Theoretical Physics, Institute of Physics, Budapest University of Technology and Economics, H-111 Budapest, Hungary}
\address{7 Center for Computational Nanoscience, A-1010 Wien, Austria}

\date{\today}

\begin{abstract}
Thin films of Co/Ru(0001) are known to exhibit an unusual spin reorientation
transition (SRT) coupled to the completion of Co atomic layers for
Co thicknesses under 4 layers. By means of spin-polarized low-energy
electron microscopy, we follow in real space the magnetization orientation
during the growth of atomically thick capping layers on Co/Ru(0001).
Capping with coinage-metal (Cu, Ag, Au) elements modifies the SRT
depending on the Co and overlayer thickness and on the overlayer material,
resulting in an expanded range of structures with high perpendicular
magnetic anisotropy. The origin of the SRT can be explained in terms of ab-initio
calculations of the layer-resolved contributions to the magnetic anisotropy
energy. Besides the changes in the SRT introduced by the capping,
a quantitative enhancement of the magnetic anisotropy is identified.
A detailed analysis of the interplay between strain and purely electronic
effects allows us to identify the conditions that lead to a high perpendicular
magnetic anisotropy in thin hcp Co films. 
\end{abstract}

\section{Introduction}

The magnetism of ultra-thin films is a fascinating field with important
device applications\cite{bookStohr}. One remarkable effect is the
film-thickness dependence of the magnetic anisotropy (MA), and particularly the
possibility of perpendicular magnetic anisotropy (PMA) in films that
are a few monolayers (ML) thick\cite{Gradmann1973}. The magnetic anisotropy
energy (MAE) responsible for this effect arises from a delicate balance
between competing contributions\cite{Sander1999,Sander2004}, including
the influence of strain in the films, as well as interactions with the
substrate. Often there is a single transition from perpendicular orientation
of the magnetic easy axis to an in-plane orientation as the magnetic
film thickness is increased. This is due to the increasing weight
of the long-range dipolar magnetostatic energy, which is reduced for
in-plane orientation of the magnetization. More unusual is the presence
of successive easy-axis reorientation transitions in thin films\cite{Panissod1992PRB,Farle1998RPP}.
In some thin film systems, the easy-axis is in-plane up to a critical
thickness, then it turns to a perpendicular orientation, and back
again to in-plane orientation at a larger thickness, i.e. they show  a double
spin reorientation transition (SRT). This is attributed to a complex
interplay of magnetic interactions influenced by atomic structure and electronic effects.
For a few systems, in particular Fe/W(110)\cite{Bergmann2004}
and Co/Ru(0001)\cite{Farid2006PRL}, it has been shown that the SRTs take place abruptly at consecutive atomic layers. 
These experimental observations can be understood
by means of ab-initio calculations that take into account epitaxial strain
as well as changes in the electronic structure of the magnetic material
that are induced by the presence of adjacent media (vacuum or substrate)\cite{Farid2006PRL}. 

For a number of reasons, it is interesting to study the effects of
capping the films with more inert, non-magnetic materials such as
gold, silver, or copper. Besides the possibility of improving the
environmental stability of magnetic transition metal films, many cases
have been observed where the addition of ultrathin layers of a non-magnetic
material can have important effects on the magnetic properties, and in particular on the MA.
 Large PMA has been obtained for a wide variety of Co
films and multilayers formed in combination with non-magnetic layers
of Pd, Pt or Au\cite{JMMM54,Velu1988,Grolier1993,Ujfalussy1996,Dorantes2003}.
In Co/Cu(100) films, the deposition of minute amounts of copper \cite{Pescia95}
can rotate within the plane the weak in-plane anisotropy of the cobalt
films. In Co/W(110) films, the addition of a Cu cap produces an increase
in the PMA at a Cu thickness close to one monolayer\cite{DudenPRB1999}.
Even adsorption of gases influences the MAE, as evident from the SRT
induced upon coverage of Co/Pt(111) films with CO \cite{Ferrer02}
and from the inverse SRT found in Fe/W \cite{Elmers1999PRB}.

Coinage-metal capping thus provides an interesting lever to control
the MA of ultrathin films. The key mechanisms that underlie the magnetic
effects induced by non-magnetic layers include crystalline structure,
strain, and electronic hybridization. First, different crystal structures
of the magnetic film and the overlayer constitute a source of strain,
and they may influence the symmetry of the lattice through changes
of the stacking sequence. This may alter the MAE \cite{Chappert1986,Dorantes1997,Park2005},
as we will discuss in detail in a forthcoming publication \cite{Silvia}.
In addition, hybridization at the interface alters the distribution
of electronic levels and subsequently the magnetic properties of both the magnetic film and the polarized cap material, with a direct impact
on the MAE.

Some Co-based thin film systems, such as those including Cu, Au and Pt\cite{Szunyogh1997,Christides1999,Hsueh2002,Park2005,Cinal2006},
have received more attention in the literature than Co/Ru \cite{Hashizume2006,Himi2001,Ding2005,Song2005,Hashizume2006,Davies2008}.
Nonetheless, Co/Ru is a particularly interesting prototypical system.
Both substrate and film material share the same hcp crystal structure, and the Ru lattice parameter is closer to Co than those of Au or Pt. Furthermore this system has a peculiar double spin-reorientation
transition, linked to the completion of atomic layers, as we showed
in a previous paper\cite{Farid2006PRL}. Building on the earlier observations
on bare Co/Ru films, we report here the changes induced in the easy-axis
of magnetization of Co/Ru(0001) films of different thicknesses as
a function of coinage-metal overlayer material and thickness. Our
results are based on measurements using spin-polarized low-energy
electron microscopy (SPLEEM) and on fully relativistic ab-initio calculations
within the screened Korringa-Kohn-Rostoker (SKKR) method. We find
that coinage-metal capping of Co/Ru films results in SRTs that depend
strongly on chemical nature as well as atomic layer thickness of the
capping layers. A summary of our measurements is shown in table \ref{summary_table}. One important difference between capped Co/Ru films, compared to the
case of bare Co/Ru films, is that the range of Co film thicknesses
for which PMA occurs is broadened, especially for the case of Au caps.
In addition, even when the capping layer does not induce changes in
the easy-axis of the magnetization of the Co film, the Curie temperature
may change. The complicated interplay of effects leading
to these results is studied by means of calculations
of the MAE which allow to separate the different contributions (strain,
hybridization, thickness) in a layer-resolved analysis. In this way we determine the factors that lead to high PMA in thin hcp Co films.

\begin{table}[htbp]
\centering
\caption{Measured easy-axis of magnetization for the different Co-film/capping-layer
combinations studied.}
\label{summary_table}
\renewcommand{\arraystretch}{1.2} 
\begin{tabular}{|c || c || c | c ||c |c ||c |c |c |}
\cline{2-9}
\multicolumn{1}{c||}{}&\multicolumn{8}{|c|}{Capping material}\\
\cline{2-9}
 \multicolumn{1}{c||}{}&\textit{Bare}&\multicolumn{2}{c||}{Ag}&\multicolumn{2}{c||}{Cu}&\multicolumn{3}{c|}{Au}\\
\cline{1-9}
 Co thickness&&\textit{1 ML}&\textit{2 ML}&\textit{1 ML}&\textit{2 ML}&\textit{1 ML}&\textit{2 ML}&\textit{3 ML}\\
\hline\cline{2-9}
2 ML &PMA&PMA&PMA&PMA&PMA&PMA&PMA&PMA\\
\cline{1-9}
3 ML &in-plane&PMA&in-plane&PMA&in-plane&PMA&PMA&PMA\\
\cline{1-9}
4 ML &in-plane&in-plane&in-plane&PMA&in-plane&PMA&PMA&PMA\\
\cline{1-9}
5 ML &in-plane&in-plane&in-plane&in-plane&in-plane&PMA&PMA&in-plane\\
\cline{1-9}
6 ML &in-plane&in-plane&in-plane&in-plane&in-plane&PMA&PMA&in-plane\\
\cline{1-9}
\end{tabular}
\end{table}

\section{Experimental details}
\label{exp_det}
The experiments were carried out in-situ in two different ultrahigh
vacuum low-energy electron microscopes (LEEM). The first one is a conventional
LEEM\cite{Bauer1994} equipped for local-area diffraction studies.
The second instrument is equipped with a spin-polarized electron gun
(SPLEEM\cite{Duden1998}), which provides magnetic contrast. Both
instruments have facilities for in-situ heating (up to 2300 K) and
cooling (down to 100 K) the samples while recording images at up to
video rate.

The Co films were grown
on two different Ru(0001) crystals, one in each experimental chamber. Both Ru
substrates were cleaned in-situ by repeated cycles of exposure to
oxygen followed by heating to 1800 K.
Both Ru substrates contained flat regions at least 100~$\mu$m wide
with mono-atomic steps separated by more than 5 $\mu$m. The metal
films (Co, Cu, Ag, Au) were grown by physical vapor deposition from
calibrated, electron-beam heated evaporators. The Co doser was charged
with a bare Co rod, while in the other dosers charges were held in
Mo-crucibles. Typical deposition rates are ~0.1-1~ML/min.

During Co growth the ruthenium crystals were heated up to between
425~K and 520~K, and the pressure remained below 4$\times10^{-10}$
torr. The growth was monitored in real time by LEEM. On large step-free
terraces, we find that Co grows in a nearly perfect layer-by-layer
mode for at least the first 8~ML. To achieve
this type of growth, it is important to avoid substrate regions with
high step density, which tend to enhance three-dimensional growth
in Co/Ru(0001) in particular\cite{FaridNJP2007}, and in strained
systems in general\cite{Ling2004a}. 

The deposition of Cu, Ag, Au capping layers was done at 513 K, 490 K, and 440 K, respectively.
The development of preparation schedules that result in atomically
perfect regions of coinage-metal capping overlayers on top of atomically
perfect regions of Co/Ru(0001) films again takes advantage of in-situ
sample growth during SPLEEM observation. It turns out that it is possible
to grow the capping layers at relatively high temperature, promoting
step-flow growth (or the formation of conveniently large islands).
At least in the case of thicker, more bulk-like Co films, the possibility
to prepare atomically sharp interfaces and capping layers with homogeneous
thickness benefits from the fact that coinage metals are immiscible
with Co in the bulk\cite{alloybook} and that coinage metals have
a lower surface energy than Co\cite{Giber1982,ChristensenPRB1997}.
Preventing alloying is more challenging in the limit of monolayer-thick
films of Co on Ru. These films are severely strained and lattice matched
to the substrate\cite{FaridNJP2007}, and alloying has been observed
in the first layer of AgCo on Ru(0001), where the chemical energy
cost of putting Ag and Co atoms in contact is overcome by the elastic
energy gain from the matching of the AgCo combination to the Ru substrate
lattice spacing\cite{Thayer2001PRL,Thayer2002PRL}. Consequently,
we have only grown Ag and Au capping layers on Co films thick enough
to be fully relaxed (at least 2 ML thick), and we did not attempt
to prepare Ag or Au caps on top of single-monolayer Co films. In case
of Cu capping layers, this type of elastic energy gain upon alloying
is not expected, because Cu is nearly lattice matched to the relaxed
Co films (mismatch is 1.5\%). Although surface-alloying of Co and
Cu has been observed in monolayer films on Ru\cite{SchmidPRL1996}
and may be unavoidable, we did explore the effects of preparing Cu
caps on all the Co films, including on unrelaxed Co monolayer films.
In case of monolayer-thick Co films capped by a single layer of Cu,
we were unable to detect a magnetic signal, indicating that these
structures are either not ferromagnetic, or have a Curie temperature
below 100 K (that temperature is the lower limit of our experimental
setup).

\begin{figure}
\centerline{\includegraphics[width=1\textwidth]{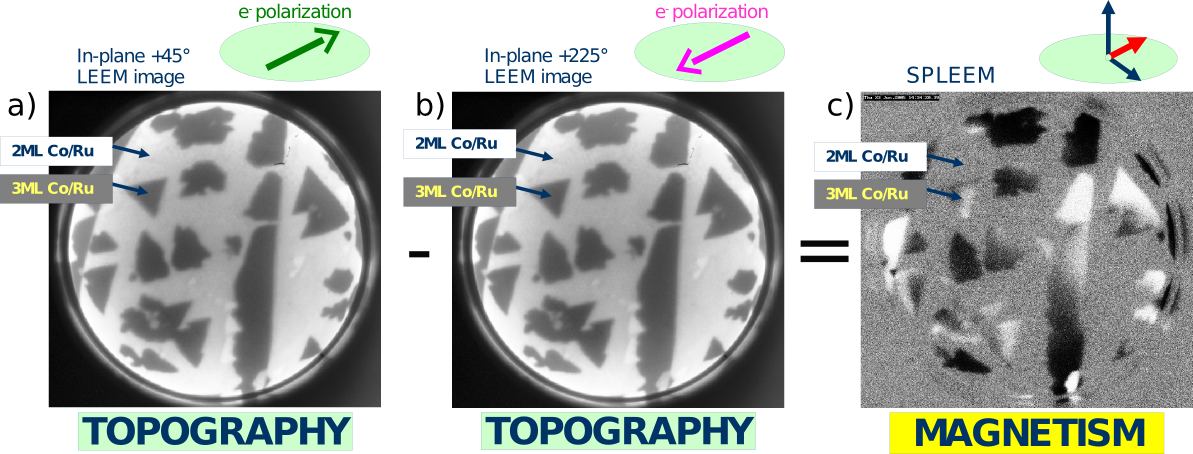}} 

\caption{Method for obtaining a SPLEEM image. A spin-polarized electron beam
is reflected off the sample surface, and a pair of images of the same
sample region is acquired. The direction of the spin-polarization
of the electron beam is rotated by 180 degrees between the two images
shown in (a) and (b). When a pixel-by-pixel difference image is formed
from these two images, all topographic, chemical, etc., image contrast
vanishes except for the contrast that is due to the sample magnetization.
Normalizing this difference image results in the grey-scale image
shown in c), where bright (dark) contrast reveals the strength of
the component of the local magnetization parallel (antiparallel) to
the direction of the spin-polarization used in a) (+45$^{\circ}$),
indicated by the red arrow in the schematic axis above panel c). Field
of view is 2.8~$\mu$m and the electron energy is 7~eV.}
\label{SPLEEM-how} 
\end{figure}

SPLEEM\cite{Duden1998} was used to monitor the easy axis of magnetization
of the films. For a spin-polarized low-energy electron beam, the reflectivity
of the sample surface depends not only on topography, chemical composition,
and other factors, but also on the relative alignment between the
beam polarization and the sample magnetization. The SPLEEM is equipped
to allow the spin direction of the electron beam to be changed to
any desired orientation\cite{Duden1995}. By acquiring pairs of images
taken with reversed spin-polarizations (Fig.~\ref{SPLEEM-how}),
we can employ pixel-by-pixel subtraction of the two images to enhance
magnetic contrast while suppressing all other forms of contrast (topography
etc.). In the resulting SPLEEM images, bright (dark) contrast indicates
that magnetization has a component parallel (antiparallel) to the
spin-up direction of the electron beam. By collecting three such pairs
of images, using three orthogonal quantization axes (usually the direction
perpendicular to the surface plus two orthogonal in-plane directions),
we can obtain triplets of SPLEEM images that reflect the 3-dimensional (3D) components
of the magnetization vector in the sample surface~\cite{Ramchal2004PRB}.

\section{Theoretical method}

Calculations have been performed within a fully relativistic ab-initio
framework based on density functional theory using the SKKR (Screened
Korringa-Kohn-Rostoker) method. The main features of this approach
are described elsewhere \cite{Weinberger2005}. Here we only mention
those relevant for the present study. Within the SKKR formalism, the
structure under study is described as a stack of layers with a common
two-dimensional (2D) lattice parameter. Consequently, the method naturally
provides the layer-resolved physical quantities. To determine the
uniaxial magnetic anisotropy of a specific structure,  we first perform
a self-consistent calculation to obtain the electronic potentials
and exchange fields, and then, applying the force theorem, we calculate the
band energy term for two orientations of the magnetization, normal
and parallel to the surface. Convergence is achieved using an energy-dependent
k-point mesh that includes as many as $4\cdot10^{4}$ points in the
irreducible Brillouin zone, so that the accuracy in the final MAE
values is 0.001 meV. The MAE is defined as the balance between the
band and dipole energy contributions:

\begin{equation}
MAE=\Delta E_{b}+\Delta E_{dd}\end{equation}

\noindent with \begin{equation}
\Delta E_{\xi}=E_{\xi}[{\bf M}_{\|}]-E_{\xi}[{\bf M}_{\perp}]\hspace*{2cm}\xi=b,dd.\end{equation}

\noindent defined as the difference between energies obtained with
the magnetization vector (${\bf M}$) contained in the surface plane
or directed along the normal to the surface. Within this convention,
a positive MAE corresponds to an easy-axis of magnetization along
the normal to the surface. The dipole energy for a particular orientation of $\bf{M}$
is obtained from the
classical interaction between magnetic dipoles in atomic Rydberg units:

\begin{equation}
E_{dd}=\frac{1}{c^{2}}\sum_{R,R'}\left\{ \frac{{\bf m}_{R}{\bf m}_{R'}}{|{\bf R}-{\bf R}'|^{3}}-3\frac{[{\bf m}_{R}\cdot({\bf R}-{\bf R}')][{\bf m}_{R'}\cdot({\bf R}-{\bf R}')]}{|{\bf R}-{\bf R}'|^{5}}\right\} \end{equation}

\noindent where ${\bf m}_{R}$ is the magnetic moment at site ${\bf R}$
and the sum is restricted to ${\bf R\neq R'}$; being a demagnetization
energy, it always favors in-plane magnetization.

The structures we have modelled are thin Co films 2-10 ML
thick on a Ru(0001) substrate, either bare or covered by a coinage-metal
capping of 1 to 10 ML thickness. To understand specific effects, other
capping metals such as Ru or Pt have also been considered. As shown
in table \ref{thtable}, there are significant structural differences
between the elements forming these structures. In our calculations,
we use a common 2D lattice parameter (a$_{2D}$) for all layers of
a given structure. In most cases, we use the intermediate value corresponding to the Ru(0001) lattice, but we also analyze the effect of different
values of a$_{2D}$ on the main results. In order to recover the atomic
volume corresponding to each element, interlayer relaxations ($\Delta$d)
were allowed. The results presented here correspond to $\Delta$d
values of -6 \% for the Co and Cu layers, and +6 \% for Au, Ag and
Pt, both with respect to the Ru interlayer distance. At the metal/Co
interfaces, the nonuniform relaxation introduced in Ref.~\cite{Farid2006PRL}
is used.

\begin{table}[htbp]
\centering 

\caption{Bulk crystal lattice type and corresponding 2D parameter (a$_{2D}$, in
$\AA$) at the fcc [111] or hcp [0001] orientations for the elements
forming the thin films under study. The last row contains the nominal
valence charge (Q) of each element.}

\label{thtable} \renewcommand{\arraystretch}{1.2}
\begin{tabular}{lllllll}
 &  &  &  &  &  & \tabularnewline
\hline 
Element  & Ru  & Co  & Cu  & Ag  & Au  & Pt \tabularnewline
\hline 
Lattice  & hcp  & hcp  & fcc  & fcc  & fcc  & fcc \tabularnewline
a$_{2D}$ & 2.71  & 2.51  & 2.55  & 2.91  & 2.87  & 2.77 \tabularnewline
Q  & 8.0  & 9.0  & 11.0  & 11.0  & 11.0  & 10.0 \tabularnewline
\end{tabular}
\end{table}

The presence of an overlayer alters the local electronic properties
of the Co film. These changes affect the layer in contact with the
capping, as well as, to a lesser extent, adjacent layers. At the bare
Co film, there is a surface-induced narrowing of the density of states
(DOS) at the topmost layer, which can still be observed (although
much reduced) at the layer below. After coverage with 1 cap layer,
this narrowing only subsists at the outermost Co plane, leading to
local band filling effects. In general, for the noble metal capping
the hybridization between Co and overlayer modifies the shape of the
Co majority spin DOS, specially for the $d$ orbitals with weight
along the normal to the surface. Table \ref{thtable2} compiles the
charge and magnetic moments at the Co/cap interface for a structure
formed by 2 cap layers on a Co film of 4 ML. These values are representative
for the rest of systems considered here. A cap Ru film behaves similarly
to the Ru substrate, at least for more than 1 ML cap, and the charge
and moments provided in the table for the Ru overlayer coincide with
those at the Co/substrate interface. Regarding the noble metal overlayers,
the charge transfer between Co and cap is not significant, contrary
to the Co/Ru interface, where Ru atoms lose 0.15 electrons. The net
magnetic polarization induced at the overlayer is negligible, even
though the distribution of electronic states is largely affected by
the hybridization with Co. On the contrary, Co attains large magnetic
moments when covered by the noble metals, with a gradual increase
of the spin moments as the spin-orbit coupling (SOC) of the overlayer
becomes important. Even higher Co spin moments can be obtained for
cap metals with an unfilled $d$ band and high SOC like Pt, where
the net magnetic moment of the structure is considerably enhanced
due to the additional polarization induced at the cap. However, there
is no correlation between the charge transfer and the induced polarization.
And, as we will show, in general neither the charge nor the magnetic
moment variations can be directly correlated to the MA.

\begin{table}[htbp]
\caption{
Charge (Q), spin moment ($m_s$) and orbital moment ($m_l$) for the atoms at the Co/cap interface 
of a Co film 4 ML thick capped by a 2ML thick overlayer. The magnetic moments are given in units of $\mu_B$.}
\label{thtable2}
\newcommand{\cq}[1]{\multicolumn{2}{c}{#1}}
\newcommand{\cqq}[1]{\multicolumn{2}{c|}{#1}}
\newcommand{\cqqq}[1]{\multicolumn{1}{c}{#1}}
\newcommand{\cc}[1]{\multicolumn{4}{c|}{#1}}
\newcommand{\cru}[1]{\multicolumn{5}{c|}{#1}}
\newcommand{\cpt}[1]{\multicolumn{3}{c}{#1}}
\newcommand{\ccc}[1]{\multicolumn{16}{c}{#1}}
\renewcommand{\tabcolsep}{0.2pc} 
\begin{tabular}{l|lr@{.}lr@{.}l|r@{.}lll|r@{.}lll|r@{.}lll|lll}
\ccc{} \\
\hline
Overlayer &  \cru{Ru} &  \cc{Cu} &  \cc{Ag} &  \cc{Au} &   \cpt{Pt} \\
           & \cqqq{Q}   &  \cq{$m_s$} & \cqq{$m_l$} & \cq{Q} & $m_s$ & $m_l$ & \cq{Q} & $m_s$ & $m_l$ & \cq{Q} & $m_s$ & $m_l$ & \cqqq{Q} & $m_s$ & $m_l$ \\
\hline
Co  &  9.12 & 1&55 & 0&11 & 8&97 & 1.71 & 0.12 & 9&03 & 1.76 & 0.12 & 9&06 & 1.81 & 0.11 & 9.03 & 1.94 & 0.11 \\
Cap &  7.84 & -0&01 & -0&00 & 11&02 & 0.02 & 0.00 & 10&96 & 0.00 & 0.00 & 10&93 & 0.01 & 0.01 & 9.93 & 0.28 & 0.06 \\
\end{tabular}
\end{table}

In the following we will concentrate on the MAE of the structures
formed by covering Co films of various thicknesses with different
noble-metal overlayers.

\section{Experimental results}

In strained systems, layer-by-layer thin film growth is unstable towards
the formation of 3D islands that can more efficiently relieve
the lattice mismatch with the substrate. When one is interested in
the precise thickness-dependence of magnetic film properties, 3D islanding
must be suppressed. In other work, the approach has often been to
deposit films at relatively low substrate temperature, where high
nucleation density can be exploited to stabilize layer-by-layer epitaxial
growth. Atomic-level film-thickness control has often been achieved
in this way. However, the film surfaces resulting from such growth
usually contain a high density of atomic steps. Thus, the thickness
of extended regions of such films is usually an average quantity,
in the sense that such films are mosaics of small regions with thicknesses
that deviate from the average value by one or more ML. 

With the goal to study magnetic properties of precisely thickness-controlled
films, we used a different approach to suppress 3D islanding tendencies.
We have found that, under the appropriate growth conditions, layer-by-layer
growth can proceed to relatively thick films (tens of layers), even
when lattice mismatch is in the range of 5-7\% \cite{Ling2004a},
even during growth at relatively high substrate temperature. Our preferred
way to suppress 3D islanding is to deposit the film material on very
large, atomically flat terraces. On atomically flat regions, the formation
of next-layer islands due to spill-over effects on downward substrate
steps is avoided and, as a result, layer-by-layer growth is extended
to greater film thicknesses than one would observe on rougher substrates.
In this way, we prepare well-annealed films that have homogeneous
thickness and no atomic steps across regions that are large enough
to be resolved and analyzed individually in our experiments. The magnification
range and fast image acquisition of low-energy electron microscopy
allows us to rapidly scan large areas of the substrates, in order
to locate appropriate atomically flat terraces, and to zoom in and
analyze homogeneous regions of the films.

Using this method to prepare and analyze regions of essentially atomically
perfect Co/Ru(0001) films, we previously found\cite{Farid2006PRL}
that only those films and islands with thickness of exactly two atomic
layers have a perpendicular easy axis of magnetization. All islands
or films with other thicknesses, i.e., single-layer films and films
with three or more layers, have an in-plane easy axis of magnetization
(we have extended the measurements to include all thicknesses up to
8~ML).

\begin{figure}
\centerline{\includegraphics[width=0.8\textwidth]{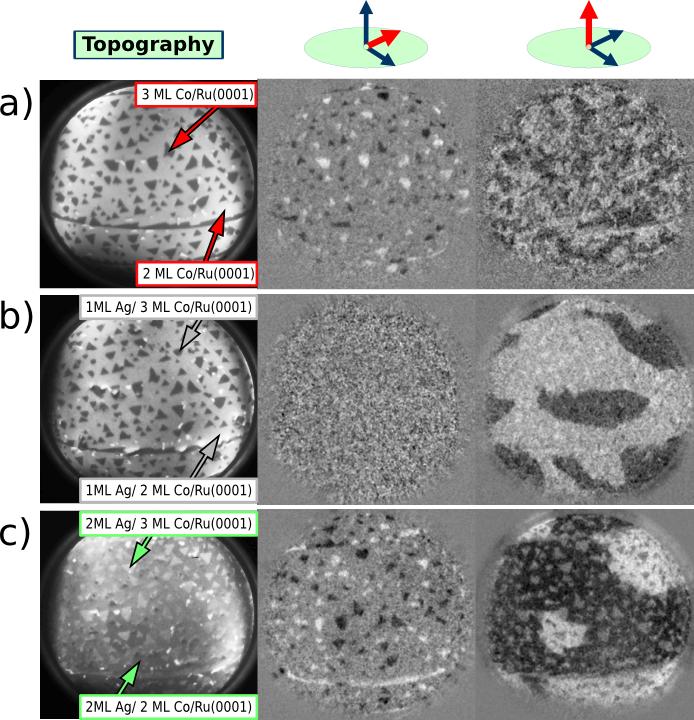}} 

\caption{LEEM images series of the topography (left column) and SPLEEM images
of the magnetic contrast in-plane (middle column) and perpendicular
to the surface(right column), of: a) 2 ML thick continuous Co film
decorated with 3 ML thick Co islands on Ru(0001), b) capped with 1~ML
of Ag, and c) capped with 2~ML of Ag. The 2~ML Co/Ru(0001) film
is magnetized out-of-plane while the 3 ML Co islands are magnetized
in-plane. The addition of the 1 ML Ag cap affects only the 3 ML Co
islands, changing their easy-axis from in-plane to out-of-plane. An
additional Ag layer (cap layer of 2 ML total thickness) changes the
3 ML islands back to an in-plane easy-axis. In contrast, in the 2
ML thick Co film we find only out-of-plane magnetized domains, independently
of the presence of cap layers. The field of view of all the images
is 7~$\mu$m, and the electron energy is 6.8, 6.0 and 6.8~eV for
images a), b) and c) respectively.}

\label{AgCoRu-3D} 
\end{figure}

\begin{figure}
\centerline{\includegraphics[height=0.8\textwidth]{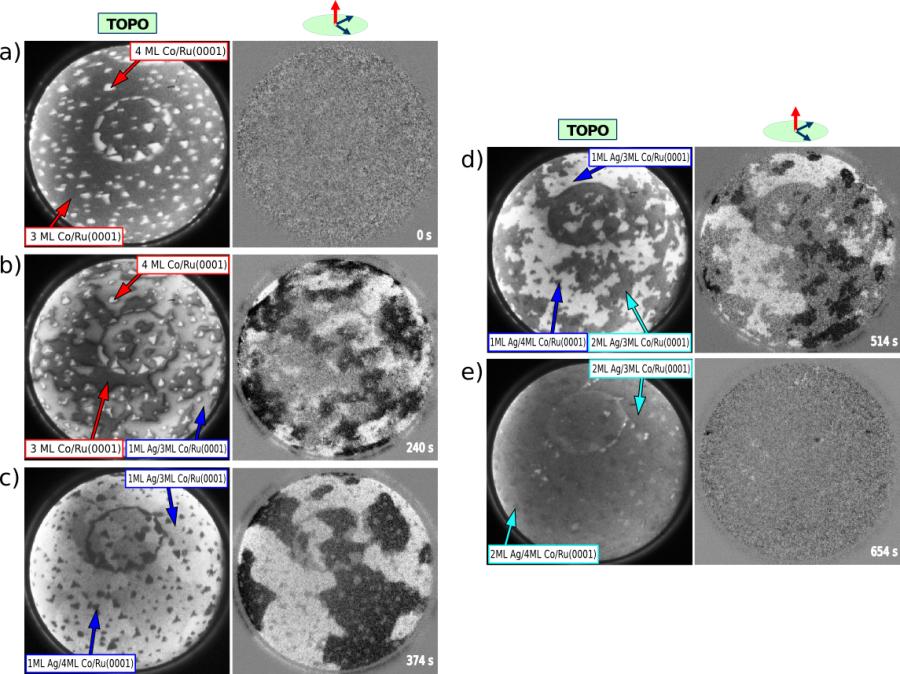}} 

\caption{LEEM images (left column) and SPLEEM images (right column, perpendicular
spin-polarization) from a movie that monitors in real time the growth
of two consecutive atomic layers of Ag on a 3 ML thick continuous Co
film decorated with 4 ML thick Co islands on Ru(0001). The sample temperature is  470~K. In the bare Co
film, (a), null-contrast in the SPLEEM image shows that both 3ML and
4 ML regions are magnetized within the plane. The first Ag monolayer,
indicated in (b, c, d), changes the easy axis only of the 3~ML thick
Co areas from in-plane to out-of-plane, as is evident from contrast
in the SPLEEM images. An additional layer of Ag, indicated in (d,
e), has the inverse effect on the 3~ML Co film, inducing a change
from perpendicular to in-plane. The Ag coverage and deposition time is marked in
the figures. The field of view is 7~$\mu$m and the electron energy
is 7.6~eV.}

\label{AgCoRu} 
\end{figure}

\begin{figure}
\centerline{\includegraphics[height=0.8\textwidth]{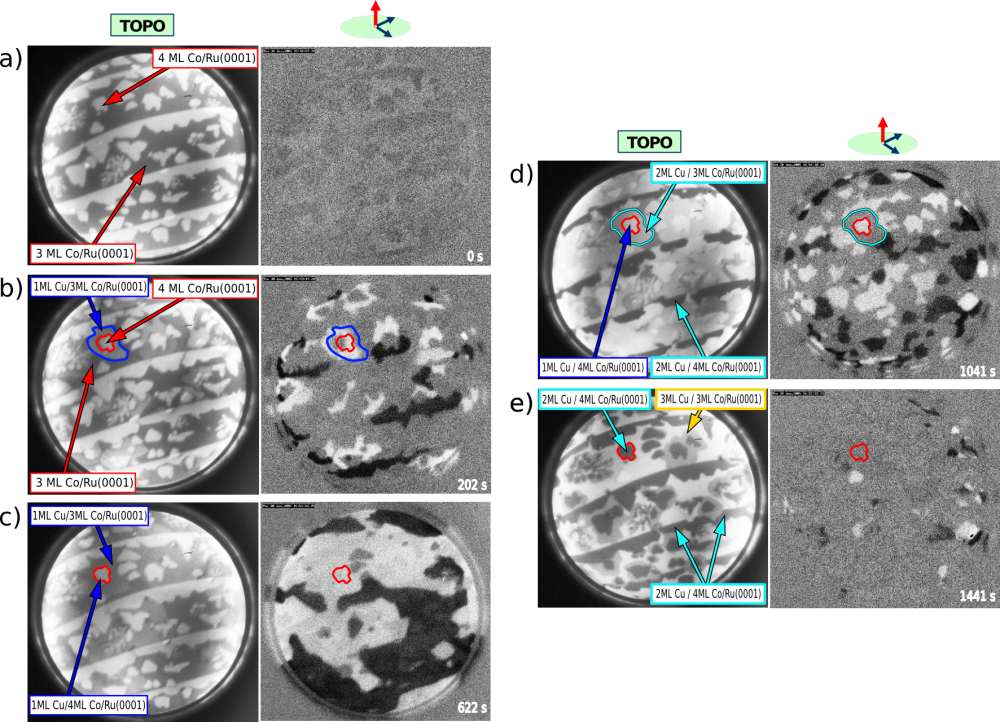}} 

\caption{LEEM images (left column) and SPLEEM images (right column, perpendicular
spin-polarization) from a movie that monitors in real time the growth
of two consecutive atomic layers of Cu on a Co continuous film 3 ML
thick decorated with additional 4 ML Co islands. The sample temperature
was 513 K. In the bare Co film, (a), null-contrast in the SPLEEM image
shows that both 3ML and 4 ML regions are magnetized within the plane.
The first Cu monolayer, (b, c) changes the easy axis of 3 ML film
and 4 ML islands from in-plane to out-of-plane. Deposition of an additional
Cu layer, (d, e) changes the magnetization back to an in-plane orientation
for both the 3 ML thick areas and the 4 ML thick islands. The Cu coverage and deposition
time is indicated. The field of view is 7~$\mu$m and the electron
energy is 7~eV.}

\label{CuCoRu} 
\end{figure}

Depositing capping layers on top of the Co/Ru films, we find that
for all combinations of overlayer metal (Ag, Au, or Cu) and Co-film thickness, growth conditions can be adjusted such that the overlayer metals grow in layer-by-layer mode (excluding the cases Ag or Au on single-monolayer Co/Ru(0001), for the reasons given in section \ref{exp_det}). Examples of this are seen in Figures~\ref{AgCoRu-3D},\ref{AgCoRu}, and \ref{CuCoRu} . The capping
overlayers start covering first the lower Co level, indicating that
the Ehrlich-Schwoebel barrier is not large enough to prevent the downhill
migration of the adatoms deposited on the 4 ML islands. Only when
the lower level is filled up, then the tops of the preexisting Co
islands are covered with the capping layer. In 2~ML Co films, the
easy-axis orientation of the magnetization remains unchanged, perpendicular
to the surface, when one or more Ag monoatomic layers are deposited
onto the Co films. The Ag capping layers do appear to lead to an increase
in the Curie temperature of the Co films. Although no attempt was
made to measure the Curie temperature carefully, we observe that magnetic
contrast disappears in bare 2 ML Co films when the sample temperature
is raised above 475 K, while the capped films show strong magnetic
contrast even at 525 K. Also deposition of capping layers of Cu or
Au on top of 2 ML Co films does not change the perpendicular easy
axis of magnetization. These observations indicate that the PMA of
cobalt bilayer films, capped or not, is quite robust. 

More dramatic effects are observed when we deposit capping layers
on top of three monolayer thick Co/Ru(0001) films. We had previously
reported\cite{Farid2006PRL} how ab-initio calculations show that
the in-plane anisotropy of these films is rather small, 0.04~mJ/m$^{2}$.
Indeed, we find that deposition of a single monolayer of any of the
coinage metals Ag, Au, or Cu on top of 3 ML Co/Ru(0001) results in
an SRT. This effect is demonstrated in experiments summarized in Fig.~\ref{AgCoRu}
and Fig.~\ref{CuCoRu}, where Ag and Cu were deposited, respectively,
on top of Co films with regions of 3 ML and 4 ML thickness. Simultaneous
SPLEEM imaging with perpendicular magnetization sensitivity (i.e.,
with the spin-polarization of the electron beam aligned in the
direction perpendicular to the sample surface) during the deposition of the capping
layers shows how any out-of-plane component of the magnetization is
absent in the bare films, whereas areas covered with a monolayer of
Cu or Ag produce strong magnetic contrast, as seen in panels c) and
d) of Fig.~\ref{AgCoRu} and Fig.~\ref{CuCoRu} (see also the on-line
full movies from which the frames of the figures have been extracted).
Similarly, single-monolayer Au caps on 3 ML Co films result in PMA
(no images shown here).

When thicker capping layers are deposited on the 3 ML Co films, then
the different capping materials lead to qualitatively different results.
While 2 ML thick Au cap layers still maintain PMA, bilayer capping
layers of either Cu or Ag trigger a second SRT, resulting in an in-plane
easy axis of magnetization. This behavior is seen in panels d) and
e) of Fig.~\ref{AgCoRu} and Fig.~\ref{CuCoRu} for Ag and Cu, respectively.
Quantitative increase of perpendicular anisotropy, as a consequence
of non-magnetic capping layers has been reported before, for example
for Cu on Co/W films\,\cite{DudenPRB1999}. However, our observations
of complete reorientation transitions, induced at the monolayer level
by non-magnetic capping layers, seem striking to us. This type of
consecutive spin-reorientation transitions is reminiscent of the transitions
that occurs for bare Co films when changing the Co thickness from
one, to two, and to three atomic layers \cite{Farid2006PRL}. 

\begin{figure}
\centerline{\includegraphics[width=0.85\textwidth]{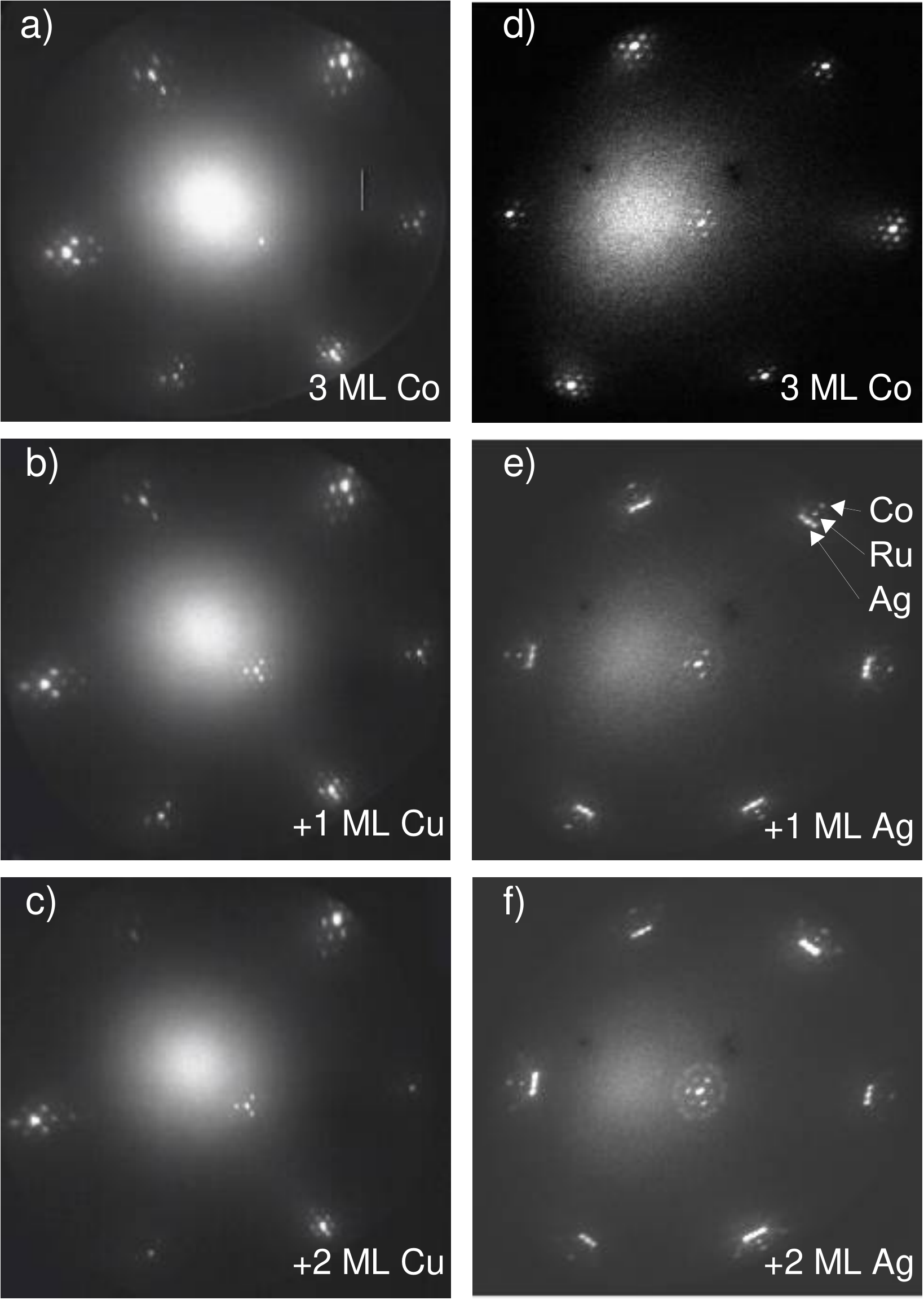}} 
\caption{LEED of bare 3ML films (a,d), and the film covered with 1 and 2 ML
of Cu (b-c) and Ag(e-f), respectively. In particular note that there
is no change for Cu, within our experimental resolution, in the spot
positions when an additional capping layer is grown on top of the
first. This is in contrast to the Ag capping layers -- labeled arrows in panel e) attribute different satellite spots to Ag and Co,
indicating that each material keeps its own lattice spacing (see text). The electron
energies are 53 and 40~eV for images a,b,c and d,e,f respectively.}
\label{LEED} 
\end{figure}

The consecutive SRTs in bare Co films \cite{Farid2006PRL} are associated
with an abrupt change in lattice spacing from the monolayer films
to the thicker films. In order to investigate the role of strain in
our capped films, we used low-energy electron diffraction (LEED).
In Fig.~\ref{LEED} LEED patterns from 3 ML Co films are reproduced,
both with and without Cu and Ag cap layers. The LEED patterns have
been acquired in-situ with the low-energy electron microscope\cite{RuSS2006}.
As seen in Fig.~\ref{LEED}(a,b), the diffraction pattern of bare
3 ML Co films have several satellite beams around each integer beam.
These patterns can be understood as moir\'{e} patterns produced by the
superposition of the relaxed, bulk-like Co lattice on the underlying
Ru lattice. Depositing 1 or 2 Cu layers on these Co films does not
produce significant changes in the diffraction patterns {[}compare
Fig.~\ref{LEED}(a) to (b,c)]. This implies that the in-plane lattice
spacing of the Cu layers is the same as that of the bare Co films,
within our error limits (we estimate error limits of the order of
2\%, mostly due to the distortions produced by the imaging optics).
This observation is consistent with the idea that, as a result of
the small lattice mismatch between bulk Co and Cu (close to 1.5\%),
the strain state of the Co films remains almost unchanged when Cu
capping layers are deposited. 

In contrast, the lattice mismatch between Co and Ag or Au is large,
over 13.6\% (Ag and Au have very similar lattice parameters). The
magnitude of the mismatch and the fact that the stress is compressive
(Ag and Au are larger than Co) suggest that monolayer cap films of
Ag or Au on top of Co/Ru(0001) films are likely relaxed. In fact,
Ag is known to relax partially when grown directly on Ru, by the introduction
of networks of misfit dislocations\cite{AgSSKevin}. We interpret
the diffraction patterns found with Ag cap layers {[}Fig.~\ref{LEED}(e,f)]
as a superposition of spots corresponding to Ag, Co and Ru {[}see
Fig.~\ref{LEED}(e)]. Within the error limit, the separation of spots
corresponds to the difference of bulk in-plane lattice parameters
of the three metals.

One can immediately appreciate from LEED patterns presented here that the forces that modify the MAE in our capped Co films must include additional factors beyond epitaxial strain. 3 ML Co capped with two monolayers of either Ag or Cu on  are magnetized in the same direction
(in-plane), even though the lattice spacings inferred from LEED for
each capping material differ greatly. On the other hand, the effect
of 2 ML Au capping 3 ML Co films is different from Ag, even though the bulk lattice spacings
of both capping materials are quite close.

The effect of capping layers on the magnetism of Co films with 4~ML
thickness is again richly dependent upon chemical nature and thickness
of the cap layer. Ag has the weakest effect on the MA of the Co films,
as the in-plane easy-axis of magnetization of the 4 ML Co films remains
stable in case of Ag capping layers of any thickness. In the case
of Cu, a single cap layer results in PMA while Cu bilayer caps (or thicker
films) return the Co magnetization to an in-plane orientation. Au
capping layers most strongly modify the MA of 4 ML Co films, 1 --
3 ML Au caps all result in PMA. 

When Co films with 5 or 6~ML thickness are capped, only Au affects
the MA sufficiently strongly to cause SRTs: 1 -- 2 ML Au capping layers
result in PMA, and for thicker Au caps the magnetization returns to
in-plane. Capping with Ag or Cu fails to produce any change in the
easy-axis of magnetization of 5 -- 6 ML cobalt films, which remain
magnetized in-plane (as the bare 5 -- 6 ML films). Finally, we have
measured the effect of cap layers on Co films 7 and 8 ML thick. At
this Co thickness range, even Au capping fails to produce PMA at any thickness. 

The summary of all the observed easy-axes in the different combinations
of magnetic film and overlayer material and thickness is shown in table \ref{summary_table}. What's most striking is the observation that
capping layers made of the nominally non-magnetic metals silver, copper,
and especially gold, appear to enhance perpendicular magnetic anisotropy
in Co/Ru(0001)-based structures. In the following section, we discuss
how this effect can be understood on the basis of ab-initio theory.

\section{Theoretical results}

The purpose of the calculations is not only
to explain the origin and provide quantitative estimates of the MAE, but to define trends with respect to its complex dependence
on the different electronic and structural conditions involved. Although
the coinage-metals are all fcc metals, their different in-plane lattice parameters a$_{2D}$
imposes distinct strain conditions in the Co film. Also the atomic number
increase from Cu to Ag to Au implies an increasing weight of spin-orbit
effects. As we will show here, both factors have a crucial impact on the MAE.
Additionally, we will demonstrate the origin of the MA dependence on both the Co and cap
films thicknesses, even though the most relevant MA effects occur at the Co/cap interface.

\subsection{Capping with 1 ML}

\begin{figure}
\centerline{\includegraphics[width=0.85\textwidth]{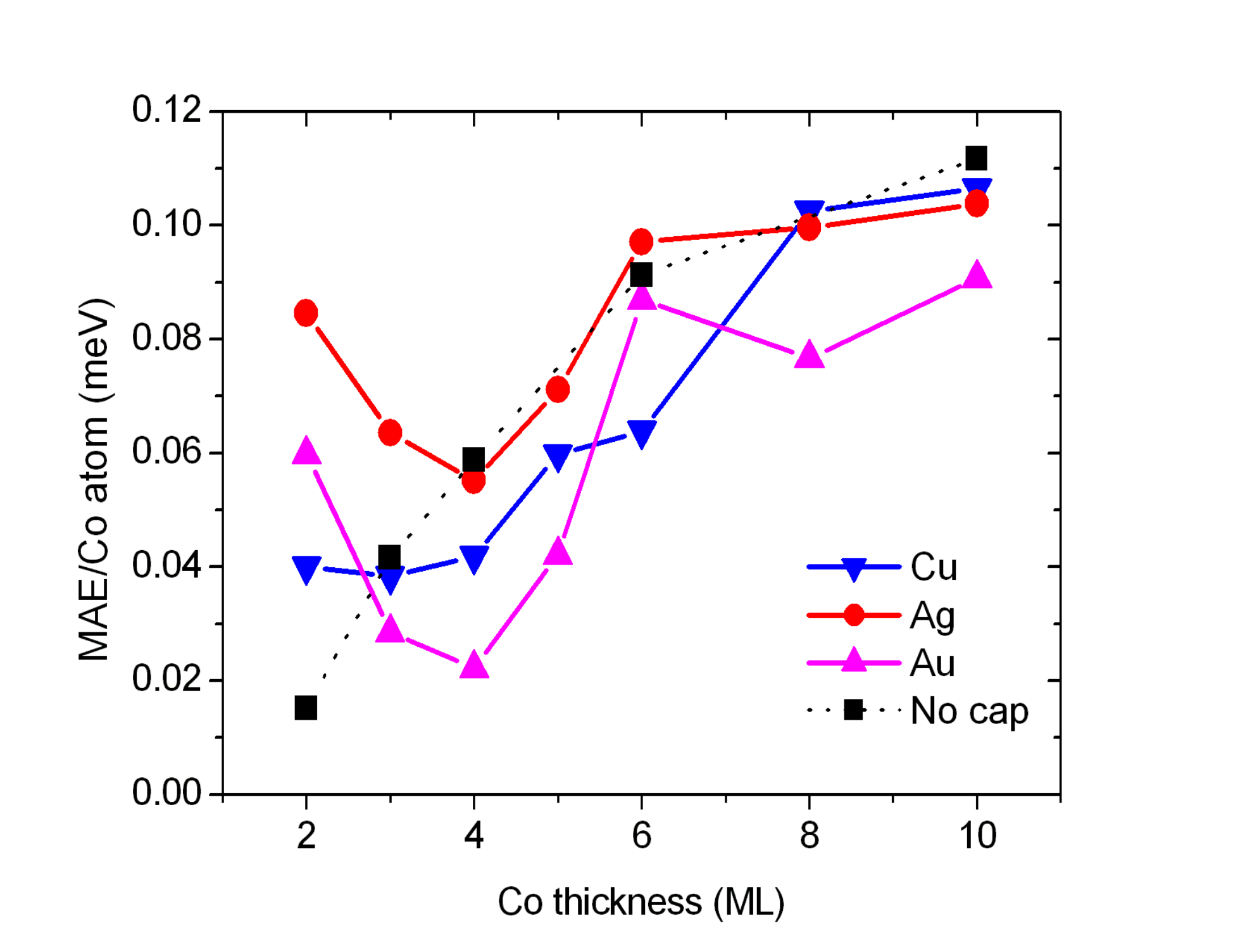}} 

\caption{MAE per Co atom for Co films of different thickness both bare and
covered by 1 cap layer of either Cu, Ag or Au.}

\label{fth-X} 
\end{figure}

We start by considering Co films of different thicknesses covered by a single coinage metal
cap layer. The summary of our results for the MAE of such structures using the a$_{2D}$
of Ru is in figure \ref{fth-X}. Notice that in the figure all MAE values
are positive, indicating an easy-axis of magnetization along the normal
to the surface. The SRT is recovered when a more realistic a$_{2D}$
closer to the Co lattice is used for the thickest films. We will come
to this point later. In order to compare the different
structures, the MAE has been normalized to the number of Co atoms,
which being the magnetic component provides the major contribution.
However, the measurements probe the MAE of the entire
film, which in the figure would amount to 1 meV for systems with 10 ML
of Co. The dependence of the MAE on the Co thickness is governed by
the $\Delta$E$_{b}$ term, as the normalized $\Delta$E$_{dd}$ is
an almost constant quantity due to the similar values of the magnetic
moments and interatomic distances for a given cap element throughout
all Co thicknesses considered. 

\begin{figure}
\centerline{\includegraphics[width=1.1\textwidth]{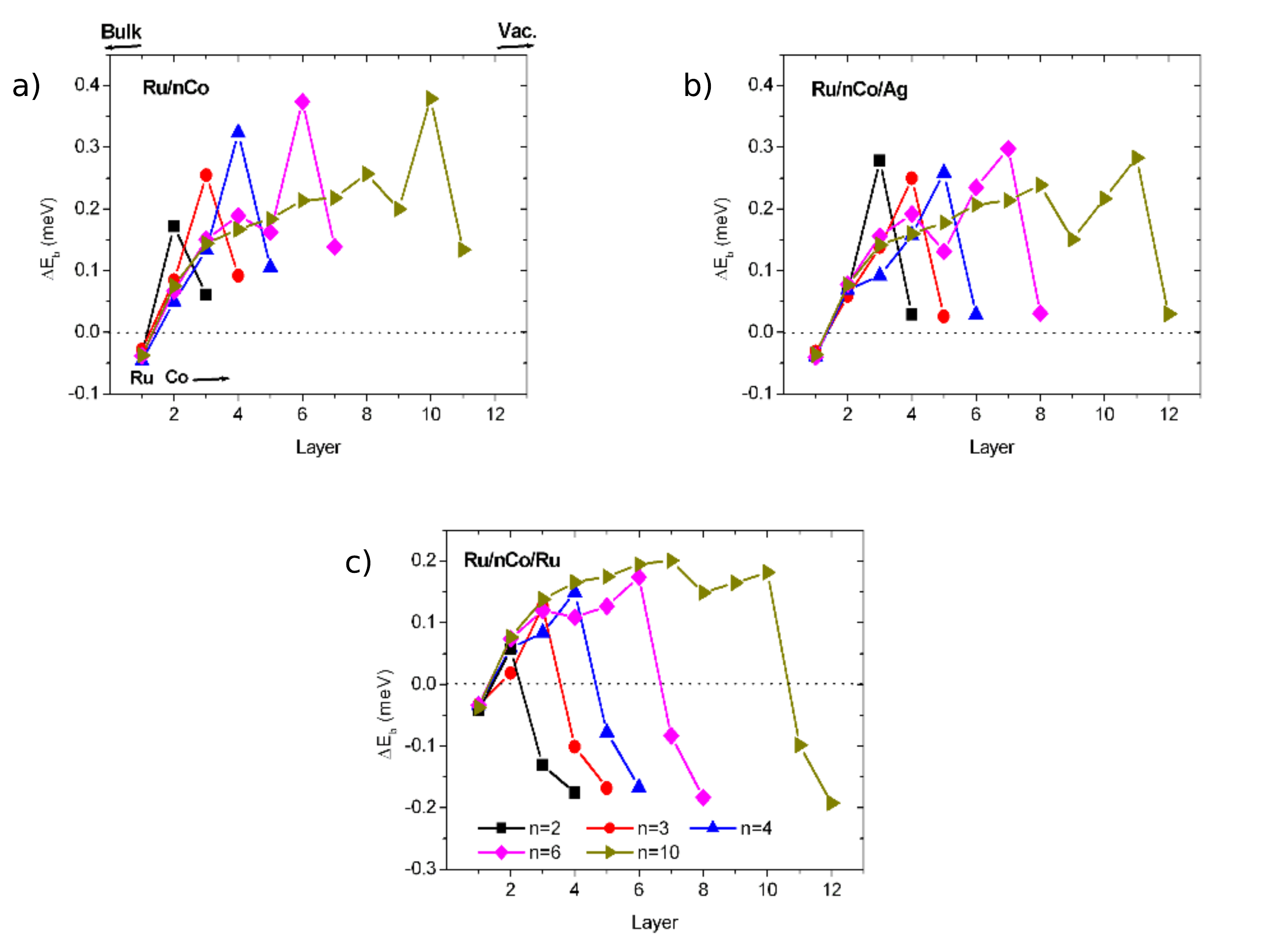}} 

\caption{Layer resolved contribution to $\Delta$E$_{b}$ for Co films of different
thicknesses (n, in ML) either a) uncapped or covered by 1 ML of b)
Ag or c) Ru. The curves of different colors at each panel correspond
to different values of n, as explained in the common legend shown
in panel c. The horizontal axis refers to layer numbers, so that each
point of a given curve provides the $\Delta$E$_{b}$ of a specific
layer. Layer numbering always starts at the substrate Ru plane closest
to the interface (layer 1), and then proceeds through the Co film
towards the surface; the highest layer number corresponds to the surface
plane, be it the outermost Co layer (panel a) or the cap layer (panels
b and c).}

\label{fth-XX} 
\end{figure}

From figure \ref{fth-X} we first note that two different thickness
regimes can be defined concerning the effect of one cap layer: for
the thinnest Co films, the MAE is considerably enhanced with respect
to bare Co, while the opposite holds for thicker Co films. Also
the differences introduced by the different cap elements are enhanced
at the thin regime. The existence of these two regimes is in
the range of the interface effects. The top panels of figure \ref{fth-XX}
provide the layer resolved $\Delta$E$_{b}$ contribution to the MAE
for the uncapped and Ag covered films. The cases with Cu and Au show
a similar layer by layer evolution as Ag. It is evident that the largest
contribution comes always from the subsurface layer. In fact the actual
value of the MAE (or of the total $\Delta$E$_{b}$ contribution)
can be viewed as a sum of two terms: a pure surface contribution,
comprising about 3 layers from the surface plane, and a contribution
from the inner layers of the Co slab. In addition, the figure proves
that two types of interfaces with opposite contributions to the MA can
be distinguished: the outermost interface with either vacuum or
a noble metal cap, and the interface with the Ru substrate. As the
range of the interface effects are similar for both, the thin regime
can be defined as Co films less than 4 layers thick, which can be
considered pure interface films. As a result, for these ultrathin
films the MAE is highly dependent on the adjacent media.

As shown in the lower panel of figure
\ref{fth-XX} for the case of a Ru cap layer, the lowering of $\Delta$E$_{b}$ at the Ru interface is not due to
the distance to the surface. Locally Ru reduces $\Delta$E$_{b}$,
even though the proximity to the surface tends to enhance the MAE,
resulting in two inequivalent Co/Ru interfaces for any selected Co
thickness. This demonstrates that surface effects must be considered
separately from the specific interactions between the materials in
contact. Consequently, in the thin film regime, interface effects are not
identical to those of a thick film or bulk-like system. The narrowing of
the DOS at the surface changes the Co/cap hybridization that ultimately
determines the surface contribution to the MAE. The complex mixing
of electronic levels induced by the spin-orbit interaction makes difficult
to provide a simple assignment of the origin of the MAE in terms of orbital levels
by visual inspection of the DOS \cite{JMMM200,Weller1995,Sawada2001,Andersson2007}.
Nevertheless, the actual changes are reflected in our layer-resolved
mapping of $\Delta$E$_{b}$, which shows that while there is a gradual
increase of the maximum $\Delta$E$_{b}$ contribution with the Co
thickness for the bare Co film, this trend disappears or significantly
reduces in the presence of a cap layer.

\begin{figure}
\centerline{\includegraphics[width=0.85\textwidth]{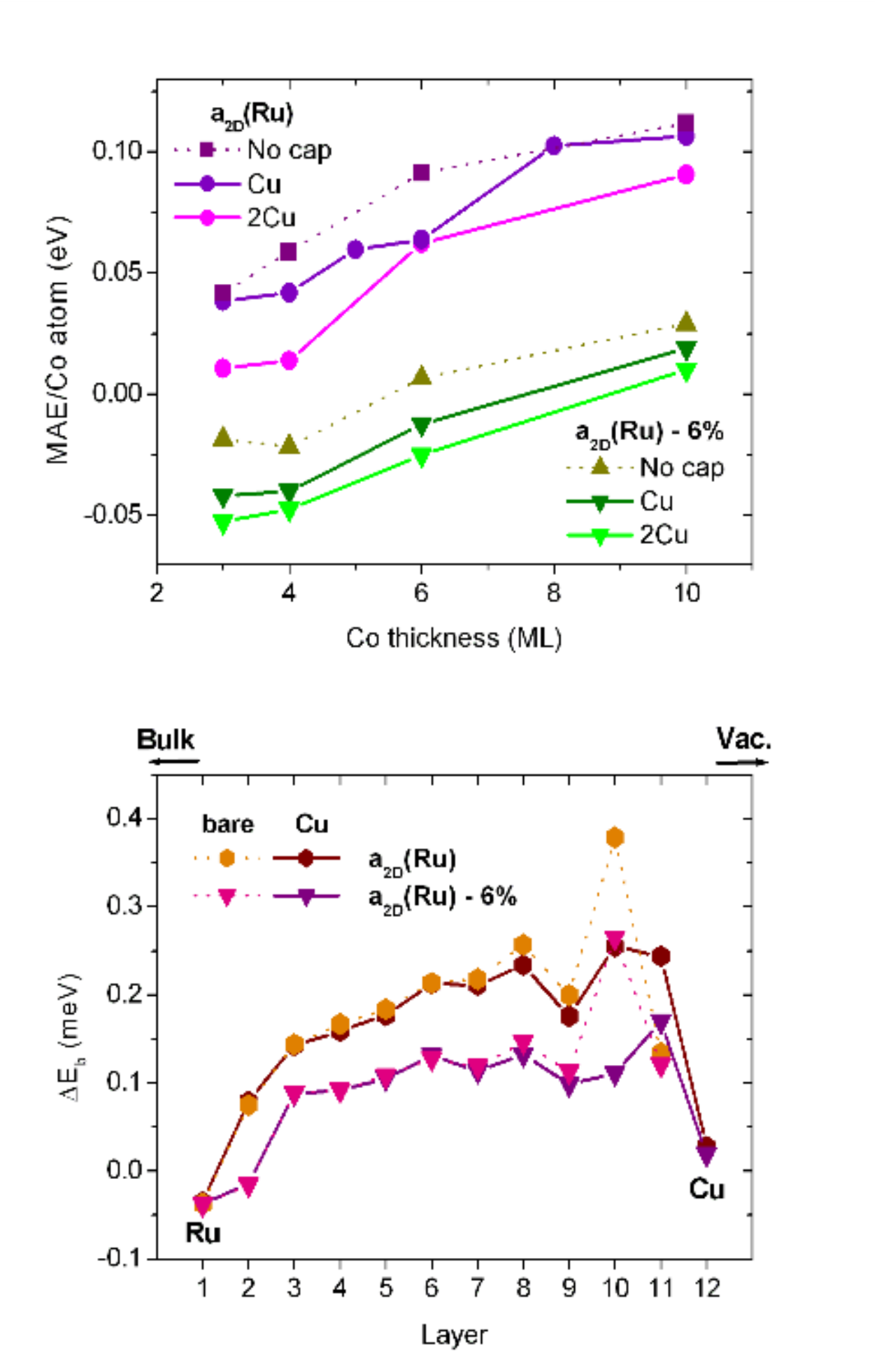}} 

\caption{(Top panel) Same as figure \protect{\ref{fth-X}} for Co films
either bare or covered by a Cu overlayer 1 or 2 ML thick, using two
different 2D lattice parameters (a$_{2D}$): that of Ru (a$_{2D}=2.71$
\AA) and a value corresponding to a 6\% of compression (a$_{2D}=2.60$
\AA). (Lower panel) Layer resolved contribution to $\Delta$E$_{b}$
for the case of a 10 ML thick Co film, comparing the bare structure
and that capped by 1 ML of Cu for both values of a$_{2D}$. The layer
numbering follows the convention of figure \protect{\ref{fth-XX}}.}

\label{fth-XXX} 
\end{figure}

In the thick film regime, the contribution of the inner Co layers to $\Delta$E$_{b}$
provides both an additive term and a background for the onset of the
surface term. In the case of a 2D expanded hcp Co lattice shown in
figures \ref{fth-X} and \ref{fth-XX}, the inner $\Delta$E$_{b}$
is high and positive, overcoming the $\Delta$E$_{dd}$ contribution
and leading to a magnetization normal to the surface. Partially
removing the strain of the Co film by reducing a$_{2D}$ towards the
Co lattice lowers the value of $\Delta$E$_{b}$ at the inner layers,
and as a result, the MAE is considerably reduced. This is shown in
figure \ref{fth-XXX} both for the bare Co film and for a Cu overlayer,
the a$_{2D}$ of Cu being similar to that of Co. The MAE becomes negative
except for the thicker Co structures, where a further 2D compression
(with the associated MAE reduction) is expected. On the other hand,
not only the layer by layer evolution of $\Delta$E$_{b}$, but also
the local effect of the capping on the surface contributions are the
same for both 2D lattices, as can be seen in the lower panel of figure
\ref{fth-XXX}. Although the strain and purely electronic effects
cannot be disentangled, the ability of our computational scheme to
separate the layer contributions helps to identify their influence on the
MAE. In fact, the relationship between the MAE and the 2D lattice
parameter evidenced here is in good agreement with the well-known
experimental evidence of large PMA for thin Co films and multilayers
on substrates with a$_{2D}\gg$ a$_{2D}^{Co}$, like Au or Pt \cite{JMMM139,Murayama1998,Wojcik2000,Chen2002,Iunin2007}.

\subsection{Thicker capping}

The enhancement of the MA at surfaces is a spin-orbit effect linked
to the surface enhancement of the spin and orbital moments, which
in turn are due to the band narrowing caused by the loss of atomic
neighbors. Intuitively one may expect that by covering a surface
with a thick capping would thus reduce the MA. This is in fact
the trend for most cap elements studied here (Cu, Ag and also Ru),
and the onset of this reduction can already be observed with 2 cap
layers (see top panel of figure \ref{fth-XXX} for the case of Cu).

\begin{figure}
\centerline{\includegraphics[width=1.0\textwidth]{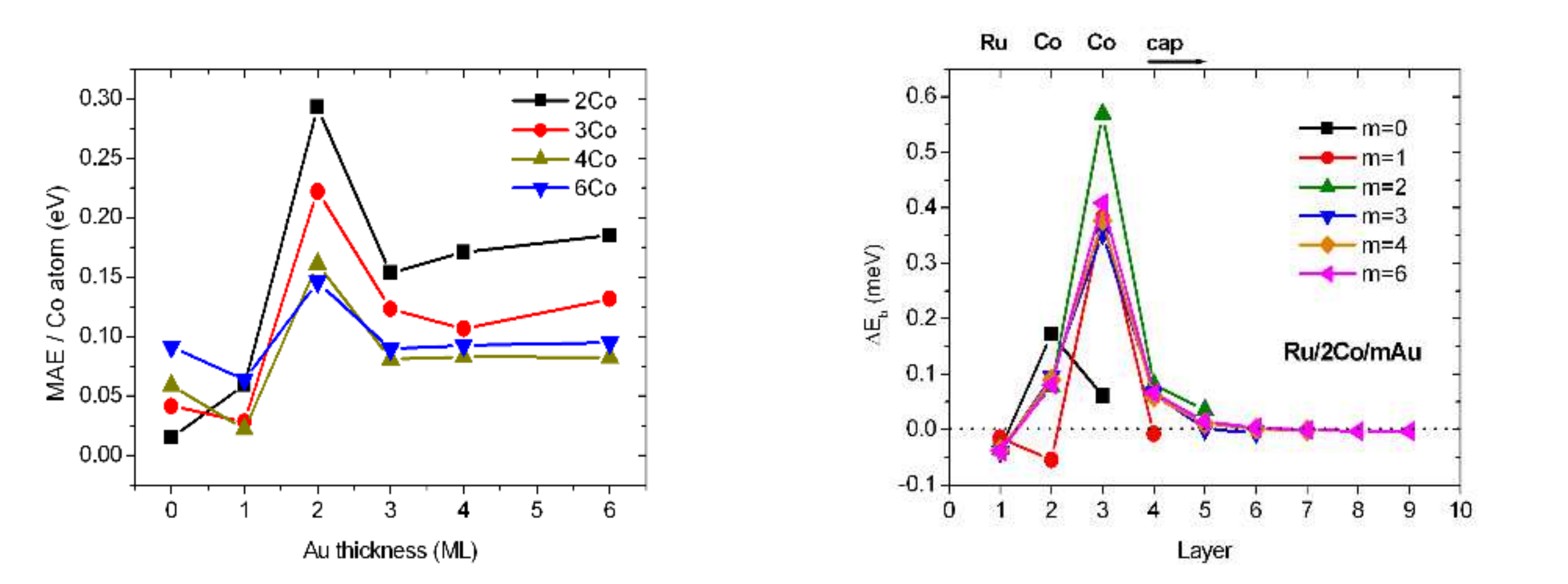}} 

\caption{(Left panel) Evolution of the MAE per Co atom with the number of capping
Au layers for Co films of different thicknesses (2, 3, 4 and 6 ML).
(Right panel) Same as figure \protect{\ref{fth-XX}} for a Co bilayer
covered with $m$ Au layers.}

\label{fth-XXXX} 
\end{figure}

However, a different situation occurs when the SOC of the cap film
becomes important, as is the case of Au. The left panel of figure
\ref{fth-XXXX} shows the evolution of the MAE for Co slabs of different
thicknesses (from 2 to 6 ML) upon thickening the Au overlayer. It
is clear that the maximum MAE per Co atom is obtained with 2 Au cap
layers for any Co thickness. In addition, thicker Au cappings always
enhance the MAE with respect to the bare Co film. This
enhancement is due to the large increase of $\Delta$E$_{b}$ at the
Co/Au interface. This contribution decreases only slightly for increasing
Au coverage. This trend is easily seen in the layer resolved contribution
shown in the right panel of the figure, corresponding to a Co thickness
of 2 ML; similar results are obtained for the thicker Co films. The
enhancement of the MAE for a bilayer capping can also be observed
for other elements with high SOC, like Pt; however, the unfilled $d$
shell of Pt favors a significant induced spin polarization, and this
influences the $\Delta$E$_{b}$ contribution of the Pt layers. Similarly
to the case of Ru, this contribution is negative for thick Pt overlayers,
and thus balances the high positive term from the Co interface.

The enhancement of the PMA for thick Au caps is in good agreement
with the SPLEEM measurements. In addition, we predict that the quantitative
value of the MAE per Co atom reaches its highest value for the combination
of Au and Co bilayers. Although the additive contribution of the layers
provides larger values of the MAE for the thicker films (for example,
for a bilayer capping of Au, the total MAE is 0.59 meV for 2
Co layers and 1.35 meV for 10 ML), for these structures the large
2D expansion of the Co film can be considered artificial. As explained in the experimental section, we expect that Au and Co tend to recover their bulk lattice parameters. As shown in figure \ref{fth-XXX}, a compression of the 2D lattice
may reduce the perpendicular MA of hcp Co films.

\begin{figure}
\centerline{\includegraphics[width=0.85\textwidth]{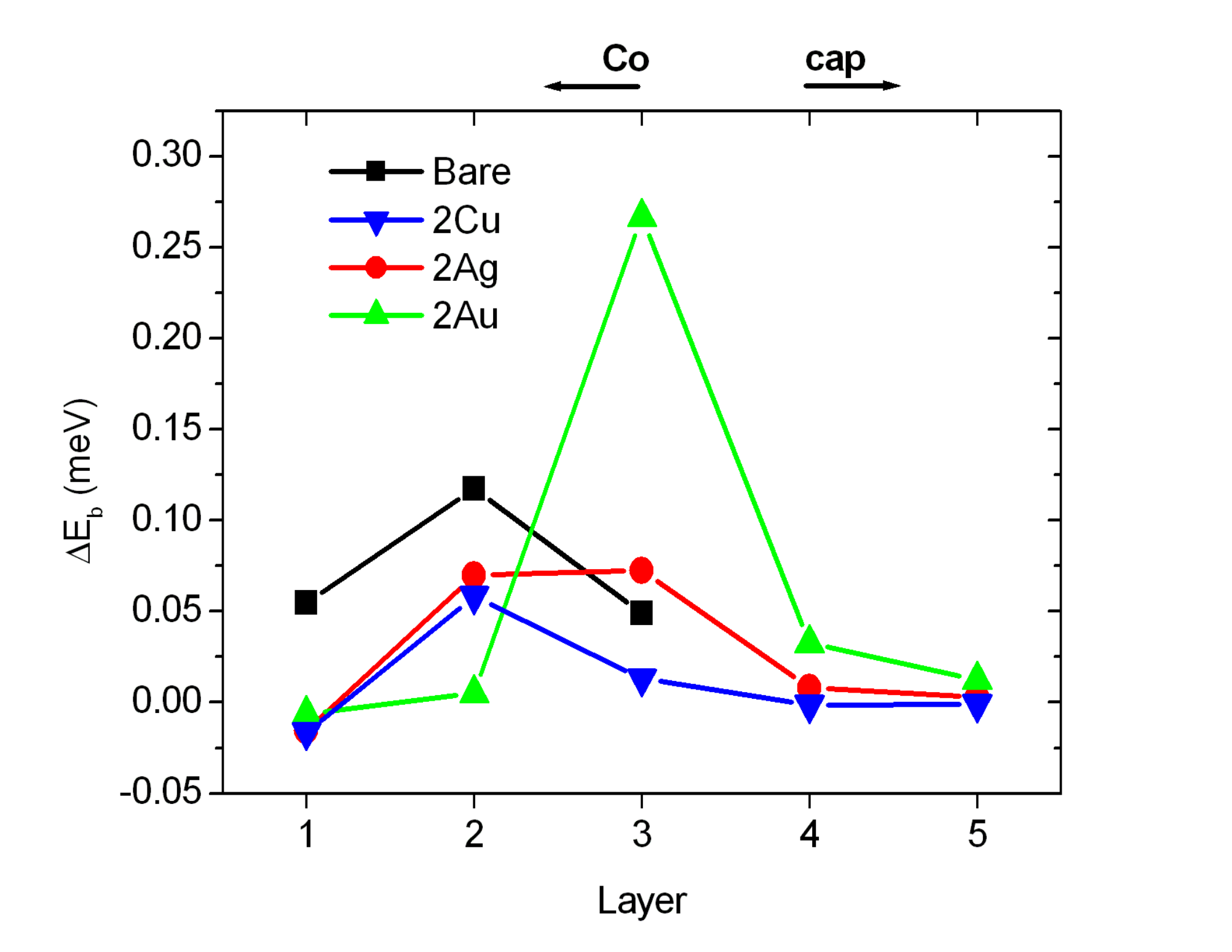}} 

\caption{Same as figure \protect\ref{fth-XX} for the Co(0001) surface of
a Co crystal either uncapped or covered by a noble metal bilayer.}

\label{fth-XXXXX} 
\end{figure}

To further explore the effect, we have modelled a semi-infinite Co(0001)
surface covered by different thicknesses (1-3 ML) of coinage-metals.
It should be kept in mind that Co is ferromagnetic so the
exchange interaction energy is several orders of magnitude larger that the
MAE. The uniaxial anisotropy is computed using a common magnetization
axis for the surface and substrate layers. The resulting easy-axis
lies in the surface plane along the {[}110] direction, in good agreement
with the experiments\cite{Pierce,PickSScom}. The local interface effect
of covering this surface with a noble metal can be seen in figure
\ref{fth-XXXXX} for a bilayer capping. Though the surface $\Delta$E$_{b}$
terms are positive, the addition of the $\Delta$E$_{dd}$ and bulk
contributions brings the easy-axis in-plane in all cases. As it occurred
for the thick Co/Ru(0001) films, capping reduces the interface
$\Delta$E$_{b}$ except for Au, where a significant enhancement occurs.
In fact, as compared to the other Au cap thicknesses, the maximum
value of $\Delta$E$_{b}$ corresponds to a Au capping of 2 ML. This
result generalizes the validity of the conclusions obtained here for
Co films on Ru(0001). The use of the intermediate a$_{2D}$ of Ru
in our calculations may be taken as representative, specially for
the dominant surface contribution. In fact, a very rough model to
approach the large lattice mismatch between the Co film and the Au
cap from the SKKR results would be to take $\Delta$E$_{b}$ of the inner layers 
from a calculation using the lattice constant of bulk Co and the surface
contribution from an expanded case. This leads to an estimate of
the SRT for a Co film capped by 2 ML Au to occur at a Co thickness
of $\sim8$ ML, in excellent agreement with the SPLEEM measurements.

\section{Summary and conclusions}

We have determined the easy-axis of magnetization of films composed
of several monolayers of Co on Ruthenium, covered with either Ag,
Cu or Au. By means of SPLEEM we have observed the changes in the easy
axis in real-time and with spatial resolution while growing the coinage
metal layers. We demonstrate the possibility of a range of structures
that have perpendicular magnetic anisotropy and Curie temperatures
well above room temperature.

The resulting MA depends at the same time on the thicknesses of both
the magnetic film and the capping overlayer, and on the element chosen
as capping metal. Co films between 3 and 6 layers in thickness present
consecutive spin-reorientation transitions coupled to the completion
of atomic layers, i.e., from in-plane magnetization to perpendicular
magnetization as the Co and/or overlayer thickness increases, and
to in-plane magnetization again for the thickest films. 
As compared to bare Co films, capping with 1 ML of Cu and Ag expand
the range of Co thicknesses for which PMA occurs. The widest range
of PMA is obtained under Au capping, where the second SRT takes place
at Co thicknesses of 7 ML for a capping of 1-2 ML of Au, or at 5 ML
for more than 2ML Au. Outside of the thickness-range where this rich
magnetic behavior is observed, Co bilayers always have perpendicular
magnetic anisotropy, irrespective of whether bare or capped with any
of the coinage metals. Similarly, coinage metal deposition on Co films
thicker than 6 layers does not affect the easy magnetization axis
(though in this case the films are magnetized in-plane).

This complex behavior can be understood in terms of the layer-resolved
contributions to the magnetic anisotropy energy. Fully relativistic
calculations based on the SKKR method allow us to identify two Co
thickness regimes defined by the range of the interface effects, which
we determine to comprise $\sim3$ layers from the interface. For ultrathin
films the MAE is governed by the dominant subsurface layer contribution,
which significantly increases upon capping by 1 ML of any coinage-metal.
At thicker films a different behavior of the surface contribution
and that from the inner layers can be identified. The first term is
reduced with respect to the thin film regime, evidencing the influence
on the MA of the proximity to the surface region. The second term
depends on the strain conditions, PMA being favored for expanded 2D
Co lattices.

The effect of the capping layer largely depends on the element chosen
as overlayer, and especially on the SOC of the cap. This is particularly
evident in the dependence of the MAE on the capping film thickness:
while thickening the Cu and Ag caps lowers the MAE, high PMA can be
obtained for Co films buried under $>6$ ML of Au, the largest anisotropy
corresponding to coverages of 2 ML.

Our results point to the wealth of possibilities to engineer the particular
easy-axis in nanometer sized structures that comes about when a precise
control of the thickness and structure of magnetic films is available.
As a rule, the ingredients to obtain a large PMA in Co films are an
expanded 2D lattice, and a thin capping with a metal of high spin-orbit
interaction. This can be best achieved with ultrathin films.

\ack This research was partly supported by the Office of Basic Energy
Sciences, Division of Materials Sciences, U.~S. Department of Energy
under Contracts No. DE-AC04-94AL85000 and No.~DE-AC02-05CH11231,
by the Spanish Ministry of Education and Science through Projects
MAT2006-13149-C02-02, MAT2006-05122, HU2006-0014 and HH2006-0027,
and by the Hungarian National Scientific Research Foundation (contract
no. OTKA T068312 and NF061726). F.E.G. and S.G. respectively acknowledge
support to the Spanish Ministry of Education and Science from an FPI fellowship
and a Ram\'{o}n y Cajal contract.

\bibliographystyle{unsrt}
\bibliography{nobleCoRu}

\begin{thebibliography}{10}

\bibitem{bookStohr}
J.~St\"{o}hr (Author) and H.C. Siegmann.
\newblock {\em Magnetism: from fundamentals to nanoscale dynamics}, volume 152
  of {\em Springer Series in Solid-State Sciences}.
\newblock Springer, Berlin, Germany, 1st edition, 2006.

\bibitem{Gradmann1973}
U.~Gradmann.
\newblock Ferromagnetism near surfaces and thin films.
\newblock {\em App. Phys.}, 3(4):161--178, 1974.

\bibitem{Sander1999}
D.~Sander.
\newblock The correlation between mechanical stress and magnetic anisotropy in
  ultrathin films.
\newblock {\em Rep. Prog. Phys.}, 62(5):809--858, 1999.

\bibitem{Sander2004}
D.~Sander.
\newblock The magnetic anisotropy and spin reorientation of nanostructures and
  nanoscale films.
\newblock {\em J. Phys.: Condensed Matter}, 16(20):R603--R636, 2004.

\bibitem{Panissod1992PRB}
K.~Ounadjela, D.~Muller, A.~Dinia, A.~Arbaoui, and P.~Panissod.
\newblock Perpendicular anisotropy and antiferromagnetic coupling in {Co}/{Ru}
  strained superlattices.
\newblock {\em Phys. Rev. B}, 45(14):7768--7771, 1992.

\bibitem{Farle1998RPP}
M.~Farle.
\newblock Ferromagnetic resonance of ultrathin metallic layers.
\newblock {\em Rep. Prog. Phys.}, 61(7):755--826, JUL 1998.

\bibitem{Bergmann2004}
Kirsten von Bergmann.
\newblock {\em Iron nanostructures studied by spin-polarised scanning tunneling
  microscopy}.
\newblock PhD thesis, Hamburg, 2004.

\bibitem{Farid2006PRL}
F.~El Gabaly, S.~Gallego, C.~Munoz, L.~Szunyogh, P.~Weinberger, C.~Klein, A.~K.
  Schmid, K.~F. McCarty, and J.~de~la Figuera.
\newblock Imaging spin reorientation transitions in consecutive atomic {Co}
  layers.
\newblock {\em Phys. Rev. Lett.}, 96:147202, 2006.

\bibitem{JMMM54}
C.~Chappert, D.~Renard, P.~Beauvillain, J.~P. Renard, and J.~Seiden.
\newblock Ferromagnetism of very thin films of nickel and cobalt.
\newblock {\em J. Mag. Mag. Mat.}, 54-57:795--796, 1986.

\bibitem{Velu1988}
E.~V\'elu, C.~Dupas, D.~Renard, J.~P. Renard, and J.~Seiden.
\newblock Enhanced magnetoresistance of ultrathin $(\frac{Au}{Co})_{n}$
  multilayers with perpendicular anisotropy.
\newblock {\em Phys. Rev. B}, 37(1):668--671, Jan 1988.

\bibitem{Grolier1993}
V.~Grolier, D.~Renard, B.~Bartenlian, P.~Beauvillain, C.~Chappert, C.~Dupas,
  J.~Ferr\'e, M.~Galtier, E.~Kolb, M.~Mulloy, J.~P. Renard, and P.~Veillet.
\newblock Unambiguous evidence of oscillatory magnetic coupling between co
  layers in ultrahigh vacuum grown {Co}/{Au}(111)/{Co} trilayers.
\newblock {\em Phys. Rev. Lett.}, 71(18):3023--3026, Nov 1993.

\bibitem{Ujfalussy1996}
B.~\'Ujfalussy, L.~Szunyogh, P.~Bruno, and P.~Weinberger.
\newblock First-principles calculation of the anomalous perpendicular
  anisotropy in a {Co} monolayer on {Au}(111).
\newblock {\em Phys. Rev. Lett.}, 77(9):1805--1808, Aug 1996.

\bibitem{Dorantes2003}
J.~Dorantes-D\'avila, H.~Dreyss\'e, and G.~M. Pastor.
\newblock Magnetic anisotropy of transition-metal interfaces from a local
  perspective: Reorientation transitions and spin-canted phases in {Pd} capped
  {Co} films on {Pd}(111).
\newblock {\em Phys. Rev. Lett.}, 91(19):197206, Nov 2003.

\bibitem{Pescia95}
W.~Weber, C.~H. Back, A.~Bischof, D.~Pescia, and R.~Allenspach.
\newblock Magnetic switching in cobalt films by adsorption of copper.
\newblock {\em Nature}, 374(6525):788--790, 1995.

\bibitem{DudenPRB1999}
T.~Duden and E.~Bauer.
\newblock Influence of {Au} and {Cu} overlayers on the magnetic structure of
  {Co} films on {W}(110).
\newblock {\em Phys. Rev. B}, 59(1):468--473, 1999.

\bibitem{Ferrer02}
O.~Robach, C.~Quiros, P.~Steadman, K.~F. Peters, E.~Lundgren, J.~Alvarez,
  H.~Isern, and S.~Ferrer.
\newblock Magnetic anisotropy of ultrathin cobalt films on {Pt}(111)
  investigated with x-ray diffraction:\quad{}effect of atomic mixing at the
  interface.
\newblock {\em Phys. Rev. B}, 65(5):054423, Jan 2002.

\bibitem{Elmers1999PRB}
H.~J. Elmers, J.~Hauschild, and U.~Gradmann.
\newblock Onset of perpendicular magnetization in nanostripe arrays of fe on
  stepped w(110) surfaces.
\newblock {\em Phys. Rev. B}, 59(5):3688--3695, 1999.

\bibitem{Chappert1986}
C.~Chappert, K.~Le Dang, P.~Beauvillain, H.~Hurdequint, and D.~Renard.
\newblock Ferromagnetic resonance studies of very thin cobalt films on a gold
  substrate.
\newblock {\em Phys. Rev. B}, 34(5):3192--3197, Sep 1986.

\bibitem{Dorantes1997}
J.~Dorantes-D\'avila, H.~Dreyss\'e, and G.~M. Pastor.
\newblock Magnetic anisotropy of close-packed (111) ultrathin transition-metal
  films:role of interlayer packing.
\newblock {\em Phys. Rev. B}, 55(22):15033--15042, Jun 1997.

\bibitem{Park2005}
S.~Park, X.~Zhang, A.~Misra, J.~D. Thompson, M.~R. Fitzsimmons, S.~Lee, and
  C.~M. Falco.
\newblock Tunable magnetic anisotropy of ultrathin {Co} layers.
\newblock {\em App. Phys. Lett.}, 86(4):042504, 2005.

\bibitem{Silvia}
S.~Gallego and M.C.~Mu\ noz.
\newblock in preparation.

\bibitem{Szunyogh1997}
L.~Szunyogh, B.~\'Ujfalussy, C.~Blaas, U.~Pustogowa, C.~Sommers, and
  P.~Weinberger.
\newblock Oscillatory behavior of the magnetic anisotropy energy in
  $cu(100)/co_{n}$ multilayer systems.
\newblock {\em Phys. Rev. B}, 56(21):14036--14044, Dec 1997.

\bibitem{Christides1999}
C.~Christides, S.~Stavroyiannis, D.~Niarchos, M.~Gioti, and S.~Logothetidis.
\newblock Dependence of the dielectric function and electronic properties on
  the co layer thickness in giant-magnetoresistance {Co}/{Au} multilayers.
\newblock {\em Phys. Rev. B}, 60(17):12239--12245, Nov 1999.

\bibitem{Hsueh2002}
H.~C. Hsueh, J.~Crain, G.~Y. Guo, H.~Y. Chen, C.~C. Lee, K.~P. Chang, and H.~L.
  Shih.
\newblock Magnetism and mechanical stability of $\alpha{}$-iron.
\newblock {\em Phys. Rev. B}, 66(5):052420, Aug 2002.

\bibitem{Cinal2006}
M.~Cinal and A.~Umerski.
\newblock Magnetic anisotropy of vicinal (001) fcc {Co} films: Role of crystal
  splitting and structure relaxation in the step-decoration effect.
\newblock {\em Phys. Rev. B}, 73(18):184423, 2006.

\bibitem{Hashizume2006}
H.~Hashizume, K.~Ishiji, J.~C. Lang, D.~Haskel, G.~Srajer, J.~Minar, and
  H.~Ebert.
\newblock Observation of x-ray magnetic circular dichroism at the {Ru} {K} edge
  in {Co}-{Ru} alloys.
\newblock {\em Phys. Rev. B}, 73(22):224416, 2006.

\bibitem{Himi2001}
K.~Himi, K.~Takanashi, S.~Mitani, M.~Yamaguchi, D.~H. Ping, K.~Hono, and
  H.~Fujimori.
\newblock Artificial modulation of magnetic structures on a monatomic layer
  scale in {Co}/{Ru} superlattices.
\newblock {\em App. Phys. Lett.}, 78(10):1436--1438, 2001.

\bibitem{Ding2005}
H.~F. Ding, A.~K. Schmid, D.~J. Keavney, Dongqi Li, R.~Cheng, J.~E. Pearson,
  F.~Y. Fradin, and S.~D. Bader.
\newblock Selective growth of {Co} nanoislands on an oxygen-patterned
  {Ru}(0001) surface.
\newblock {\em Phys. Rev. B}, 72(3):035413, 2005.

\bibitem{Song2005}
C.~Song, X.~X. Wei, K.~W. Geng, F.~Zeng, and F.~Pan.
\newblock Magnetic-moment enhancement and sharp positive magnetoresistance in
  {Co}/{Ru} multilayers.
\newblock {\em Phys. Rev. B}, 72(18):184412, 2005.

\bibitem{Davies2008}
Joseph~E. Davies, Olav Hellwig, Eric~E. Fullerton, and Kai Liu.
\newblock Temperature-dependent magnetization reversal in ({Co}/{Pt})/{Ru}
  multilayers.
\newblock {\em Phys. Rev. B}, 77(1):014421, 2008.

\bibitem{Bauer1994}
E.~Bauer.
\newblock Low energy electron microscopy.
\newblock {\em Rep. Prog. Phys.}, 57(9):895--938, 1994.

\bibitem{Duden1998}
T.~Duden and E.~Bauer.
\newblock Spin-polarized low energy electron microscopy.
\newblock {\em Surf. Rev. Lett.}, 5(6):1213--1219, 1998.

\bibitem{FaridNJP2007}
F.~El Gabaly, J.~Puerta, C.~Klein, A.~Saa, A.~Schmid, K.~McCarty, J.~Cerda, and
  J.~de~la Figuera.
\newblock Structure and morphology of ultrathin {Co}/{Ru}(0001) films.
\newblock {\em New J. Phys.}, 9, 2007.

\bibitem{Ling2004a}
W.~L. Ling, T.~Giessel, K.~Thurmer, R.~Q. Hwang, N.~C. Bartelt, and K.~F.
  McCarty.
\newblock Crucial role of substrate steps in de-wetting of crystalline thin
  films.
\newblock {\em Surf. Sci.}, 570(3):L297--L303, 2004.

\bibitem{alloybook}
T.~B. Massalski, editor.
\newblock {\em Binary Alloy Phase Diagrams}.
\newblock ASM International, Ohio, USA, 2nd edition, 1990.

\bibitem{Giber1982}
L.~Z. Mezey and J.~Giber.
\newblock The surface free-energies of solid chemical-elements -- calculation
  from internal free enthalpies of atomization.
\newblock {\em Japanese Journal of Applied Physics Part 1}, 21(11):1569--1571,
  1982.

\bibitem{ChristensenPRB1997}
A.~Christensen, A.~V. Ruban, P.~Stoltze, K.~W. Jacobsen, H.~L. Skriver, J.~K.
  Norskov, and F.~Besenbacher.
\newblock Phase diagrams for surface alloys.
\newblock {\em Phys. Rev. B}, 56(10):5822--5834, 1997.

\bibitem{Thayer2001PRL}
G.~E. Thayer, V.~Ozolins, A.~K. Schmid, N.~C. Bartelt, H.~Asta, J.~J. Hoyt,
  S.~Chiang, and R.~Q. Hwang.
\newblock Role of stress in thin film alloy thermodynamics: Competition between
  alloying and dislocation formation.
\newblock {\em Phys. Rev. Lett.}, 86(4):660--663, Mar 2001.
\newblock stress, surface alloy, Sandia.

\bibitem{Thayer2002PRL}
G.~E. Thayer, N.~C. Bartelt, V.~Ozolins, A.~K. Schmid, S.~Chiang, and R.~Q.
  Hwang.
\newblock Linking surface stress to surface structure: Measurement of atomic
  strain in a surface alloy using scanning tunneling microscopy.
\newblock {\em Phys. Rev. Lett.}, 89:036101, Jul 2002.

\bibitem{SchmidPRL1996}
A.~K. Schmid, J.~C. Hamilton, N.~C. Bartelt, and R.~Q. Hwang.
\newblock Surface alloy formation by interdiffusion across a linear interface.
\newblock {\em Phys. Rev. Lett.}, 77(14):2977--2980, 1996.

\bibitem{Duden1995}
E.~Duden, T.and~Bauer.
\newblock A compact electron-spin-polarization manipulator.
\newblock {\em Rev. Sci. Inst.}, 66(4):2861--2865, APR 1995.
\newblock spleem.

\bibitem{Ramchal2004PRB}
R.~Ramchal, A.~K. Schmid, M.~Farle, and H.~Poppa.
\newblock Spiral-like continuous spin-reorientation transition of fe/ni
  bilayers on {Cu}(100).
\newblock {\em Phys. Rev. B}, 69:214401, 2004.

\bibitem{Weinberger2005}
J.~Zabloudil, R.~Hammerling, L.~Szunyogh, and P.~Weinberger.
\newblock {\em Electron Scattering in Solid Matter: a theoretical and
  computational treatise}.
\newblock Springer-Verlag, 2005.

\bibitem{RuSS2006}
J.~de~la Figuera, F.~El Gabaly, J.~M. Puerta, J.~I. Cerda, and K.~F. McCarty.
\newblock Determining the structure of {Ru}(0001) from low-energy electron
  diffraction of a single terrace.
\newblock {\em Surf. Sci.}, 600:L105, 2006.

\bibitem{AgSSKevin}
Herringbone, {Au} triangular patterns of dislocations~in {Ag}, and {AgAu} alloy
  films~on {Ru}(0 0 0~1).
\newblock W.l. ling and j.c. hamilton and k. thürmer and g.e. thayer and j. de
  la figuera and r.q. hwang and c.b. carterb and n.c. bartelt and k.f. mccarty.
\newblock {\em Surf. Sci.}, 600:1735--1767, 2006.

\bibitem{JMMM200}
J.~St\"ohr.
\newblock Exploring the microscopic origin of magnetic anisotropies with x-ray
  magnetic circular dichroism ({XMCD}) spectroscopy.
\newblock {\em J. Mag. Mag. Mat.}, 200:470--497, 1999.

\bibitem{Weller1995}
D.~Weller, J.~St\"ohr, R.~Nakajima, A.~Carl, M.~G. Samant, C.~Chappert,
  R.~M\'egy, P.~Beauvillain, P.~Veillet, and G.~A. Held.
\newblock Microscopic origin of magnetic anisotropy in {Au}/{Co}/{Au} probed
  with x-ray magnetic circular dichroism.
\newblock {\em Phys. Rev. Lett.}, 75(20):3752--3755, Nov 1995.

\bibitem{Sawada2001}
M.~Sawada, K.~Hayashi, and A.~Kakizaki.
\newblock Electronic structure and magnetic anisotropy of
  {Co}/{Au}(111):\quad{}a spin-resolved photoelectron spectroscopy study.
\newblock {\em Phys. Rev. B}, 63(19):195407, Apr 2001.

\bibitem{Andersson2007}
C.~Andersson, B.~Sanyal, O.~Eriksson, L.~Nordstrom, O.~Karis, D.~Arvanitis,
  T.~Konishi, E.~Holub-Krappe, and J.~Hunter Dunn.
\newblock Influence of ligand states on the relationship between orbital moment
  and magnetocrystalline anisotropy.
\newblock {\em Phys. Rev. Lett.}, 99(17):177207, 2007.

\bibitem{JMMM139}
J.~Kohlhepp and U.~Gradmann.
\newblock Magnetic surface anisotropies of {Co}(0001)-based interfaces from in
  situ magnetometry of {Co} films on {Pd}(111), covered with ultrathin films of
  {Pd} and {Ag}.
\newblock {\em J. Mag. Mag. Mat.}, 139:347--354, 1995.

\bibitem{Murayama1998}
Akihiro Murayama, Kyoko Hyomi, James Eickmann, and Charles~M. Falco.
\newblock Underlayer-induced perpendicular magnetic anisotropy in ultrathin
  {Co}/{Au}/{Cu}(111) films: A spin-wave brillouin-scattering study.
\newblock {\em Phys. Rev. B}, 58(13):8596--8604, Oct 1998.

\bibitem{Wojcik2000}
M.~Wojcik, C.~Christides, E.~Jedryka, S.~Nadolski, and I.~Panagiotopoulos.
\newblock Formation of a co nanostructure revealed by $^{59}co$ nuclear
  magnetic resonance measurements in {Co}/{Au} multilayers.
\newblock {\em Phys. Rev. B}, 63(1):012102, Dec 2000.

\bibitem{Chen2002}
F.~C. Chen, Y.~E. Wu, C.~W. Su, and C.~S. Shern.
\newblock {Ag}-induced spin-reorientation transition of {Co} ultrathin films on
  {Pt}(111).
\newblock {\em Phys. Rev. B}, 66(18):184417, Nov 2002.

\bibitem{Iunin2007}
Y.~L. Iunin, Y.~P. Kabanov, V.~I. Nikitenko, X.~M. Cheng, D.~Clarke, O.~A.
  Tretiakov, O.~Tchernyshyov, A.~J. Shapiro, R.~D. Shull, and C.~L. Chien.
\newblock Asymmetric domain nucleation and unusual magnetization reversal in
  ultrathin {Co} films with perpendicular anisotropy.
\newblock {\em Phys. Rev. Lett.}, 98(11):117204, 2007.

\bibitem{Pierce}
J.~Unguris, M.~R. Scheinfein, R.~J. Celotta, and D.~T. Pierce.
\newblock Magnetic microstructure of the (0001) surface of hcp cobalt.
\newblock {\em Appl. Phys. Lett.}, 55:2553--2555, 1989.

\bibitem{PickSScom}
S.~Pick et~al.
\newblock A tight-binding study of surface magnetic anisotropy of the {Co}
  (0001) and its perturbation by {Cu} and {CO}.
\newblock {\em Sol. Stat. Comm}, 127:531--534, 2003.

\end{thebibliography}

\end{document}